\documentclass[structabstract]{aa}  
\usepackage{graphicx}
\usepackage{txfonts}
\usepackage{natbib} 
\bibpunct{(}{)}{;}{a}{}{,} 
\begin{document}
\title{Doppler imaging of the helium-variable star a\,Cen\thanks{Based on observations obtained
	  at the Canada-France-Hawaii Telescope (CFHT) which is operated by
	  the National Research Council of Canada, The Institut National des 
	  Sciences de l'Univers of the Centre National de la Recherche
	  Scientifique of France, and the University of Hawaii.}}
   
   \author{D.A. Bohlender \inst{1}
          \and
          	J.B. Rice
          \inst{2}
          \and
          	P. Hechler \inst{2}
          }

   \institute{National Research Council of Canada, Herzberg Institute of
	     Astrophysics, 5071 West Saanich Road, Victoria, BC, V9E~2E7,
	     Canada\\
             \email{david.bohlender@nrc-cnrc.gc.ca}
         \and
             Department of Physics and Astronomy, Brandon University,
              Brandon, MB, R7A~6A9, Canada \\
              \email{rice@brandonu.ca}
             }

   \date{Received 2 February 2010 / Accepted 14 June 2010}

\abstract
{}
{The helium-peculiar star a\,Cen exhibits interesting line profile variations 
of elements such as iron, nitrogen and oxygen in addition to its 
well-known extreme helium variability.  
The objective of this paper is to use new high signal-to-noise, high-resolution 
spectra to perform a quantitative measurement of the helium, iron, nitrogen and oxygen abundances of the 
star and determine the relation of the concentrations of the heavier elements on 
the surface of the star to the helium concentration and perhaps to the magnetic 
field orientation.}
{Doppler images have been created for the elements helium, iron, nitrogen and oxygen 
using the programs described in earlier papers by Rice and others.  
An alternative surface abundance mapping code has been used to model the helium 
line variations after our Doppler imaging of certain individual helium lines produced mediocre results.}
{Doppler imaging of the helium abundance of a\,Cen confirms the long-known existence of
helium-rich and helium-poor hemispheres on the star and we measure a difference of more than two orders of
magnitude in helium abundance from one side of the star to the other.
Helium is overabundant by a factor of about 5 over much of the helium-rich hemisphere.
Of particular note is our discovery that the helium-poor hemisphere has a very high abundance of
 \element[][3]{He}, approximately equal to the \element[][4]{He} abundance.  
 a\,Cen is therefore a new member of the small group of
helium-3 stars and the first well-established magnetic member of the class.
For the three metals investigated here, there are two
strong concentrations of abundance near the equator at longitude roughly 135\degr\ 
consistent with the positive magnetic maximum and two somewhat weaker concentrations 
of abundance near longitude 315\degr\ on the equator near where the helium concentration is 
centered and roughly where the negative peak of the magnetic field 
would be found.
Another strong concentration is found near the equator at about longitude 45\degr\ and 
this is not explainable in terms of any simple symmetry with the helium abundance or the
apparent magnetic field main polar locations.}
{}

\maketitle

%

\section{Introduction}

The magnetic peculiar B (Bp) star a\,Cen (HR\,5378, HD\,125823, V761 Cen) is 
arguably one of the most remarkable members of the Bp class.  
As noted by \citet{jaschek67} and \citet{jaschek68}, low-dispersion spectra 
show that the MK type of the star, determined primarily by the strengths of 
the \ion{He}{i} lines, varies between B2\,V and B8\,IV, while the strengths 
of \ion{Si}{ii} and \ion{Si}{iii} lines observed at higher dispersion remain 
essentially unchanged and correspond to a spectral type of B2.
\citet{norris68} demonstrated that at the B2 phase the helium lines are 
stronger than in a normal B2\,V standard while near B8 the star displays a 
pronounced helium deficiency and all except the strongest \ion{He}{i} lines disappear.
He also found that the \ion{He}{i} equivalent widths, W$_{\lambda}$, varied 
on a $8\fd812 \pm 0\fd004$ period.  

After the discovery of this extreme helium variability, a\,Cen was studied by 
numerous authors for its light and colour variability, line excitation anomalies 
and line strength variations.  
\citet{norris68} failed to detect a magnetic field stronger than 700\,G but a 
relatively weak field was tentatively detected by \citet{wolff74} and 
quantitatively confirmed by \cite{borra83}.
Photometric, spectroscopic and magnetic field data all vary on the \citet{norris68} 
period, assumed to be the rotation period of the star.

Qualitative discussions of the distribution of helium on the surface of a\,Cen have 
been given by \citet{norris72} and \citet{mihalas73}. 
The former proposed a single helium-strong pole on the surface of a\,Cen while the 
latter looked at a number of options including the possibility that the period was 
twice as long as the 8\fd8 period determined from the star's light curve. 
For the single wave 8\fd8 period \citet{mihalas73} explored two possibilities for 
the helium distribution: a model with two small caps of excess helium abundance with an 
inclination of the star of 45\degr\ and magnetic obliquity $\beta$ of 45\degr, and a second model with 
a single cap of excess helium with an inclination of the star of 45\degr\ and $\beta$ 
of 90\degr.

The magnetic variations of a\,Cen measured by \citet{borra83} ruled out the two-cap 
or symmetric model of \citet{mihalas73} as they found the sinusoidal longitudinal 
magnetic field curve to be nearly symmetric with extrema of -470 and +430\,G. 
Maximum \ion{He}{i} line strength occurs near the negative magnetic extremum. 
Adopting the usual Oblique Rotator Model for the star, their work then leads to the 
conclusion that our views of the negative and positive magnetic poles are very similar 
so if the inclination is not close to 90\degr, then $\beta$ must be about 90\degr. 
This then seems to be consistent with the ideas presented by \citet{norris72} and the 
single cap model of \citet{mihalas73} where a\,Cen has a single helium-strong cap 
centered approximately 90\degr\ from the rotational axis (i.e. near the negative 
magnetic pole) and the star has an inclination of about 45\degr. 

Considering the star's unique variability and its brightness ($V=4.401$), it is 
somewhat surprising that there has been essentially no detailed investigations of 
a\,Cen published since the advent of modern solid-state detectors in the 1980's 
until the recent paper of \citet{hubrig07}. 
They demonstrate that the star's metallic line spectroscopic variability is quite 
complex and note that a characteristic of a\,Cen is that some elements such as 
oxygen have lines where the neutral line strength varies in antiphase to the 
singly ionized lines.
They also found weak, variable high-excitation emission lines of \ion{Si}{ii}, 
\ion{Mn}{ii} and \ion{Fe}{ii} in the spectrum of the star but no emission in the 
H or \ion{He}{i} lines. 
They also noted the split appearence of the \ion{Fe}{ii} and \ion{Mn}{ii} 
absorption line profiles, especially at the helium line strength minimum when 
the Fe and Mn lines are at maximum strength, 
and comment upon the great difficulty of creating a Doppler image given 
the $v\sin{i}$ of the star is only about 15\,km\,s$^{-1}$.
 
Our interest in a\,Cen arose from our investigation of the stratification of 
\element[][3]{He} and \element[][4]{He} in several helium-3 stars \citep{bohlender05}, the apparent 
stratification of helium in a\,Cen \citep{leone97}, and the suggestion of 
\citet{hunger99} that a\,Cen may be an important transitional object where 
the effects of fractionation of radiatively driven winds may become important.
We obtained a tentative hint of the presence of \element[][3]{He} in a\,Cen from a single 
observation of the \ion{He}{i} $\lambda$6678 line obtained during the course 
of the helium-3 star investigation and therefore in 2003 we subsequently obtained 
observations of several \ion{He}{i} lines throughout the 8\fd8 rotation 
period of the star.  
Along with the remarkable \ion{He}{i} line variations it was immediately 
obvious that the star showed pronounced variations of metal lines as later 
reported by \citet{hubrig07}.  
Even though the $v\sin{i}$ of a\,Cen is at the extreme low end of what is 
useful for Doppler imaging and our phase resolution is quite coarse, the strong spectrum variability encouraged us to try to map the surface 
abundance distribution for the elements He, Fe, N, and O to see if 
there were indications of axial symmetry in these distributions that would 
augment the impressions of the orientation of the magnetic axis gleaned from 
the earlier work of \citet{borra83}.  
 

\begin{table*}[!t]
\caption{Spectroscopic observation log for a\,Cen}
\label{slog}
\centering
{
 \begin{tabular}{ccrccccrc}
  \hline\hline
 \multicolumn{2}{c}{4412 - 4488\AA} & \multicolumn{2}{c}{R = 93,000} & & \multicolumn{2}{c}{7238-7321\AA} & \multicolumn{2}{c}{R = 158,000} \\
 \hline
 HJD & \multicolumn{2}{c}{Phase} & Exp. time & & HJD & \multicolumn{2}{c}{Phase} & Exp. time  \\
 (2450000+) & (Decimal) & (\degr)~~ & (s) & & (2450000+) & (Decimal) & (\degr)~~ & (s)   \\
  \hline   
  2801.8505 & 0.337 & 121.3 & 600 & & 2800.7844  & 0.216 & 77.6 & 600 \\
  2801.8579 & 0.338 & 121.6 & 600 & & 2801.7660  & 0.327 & 117.6 & 600 \\
  2802.8581 & 0.451 & 162.4 & 600 & & 2802.7766  & 0.441 & 158.9 & 600 \\ 
  2802.8655 & 0.452 & 162.7 & 600 & & 2803.7557  & 0.553 & 198.9 & 600\\
  2803.8453 & 0.563 & 202.7 & 600 & & 2804.7718  & 0.668 & 240.4 & 600\\
  2803.8529 & 0.564 & 203.0 & 600 & & 2805.7523  & 0.779 & 280.4 & 600 \\
  2804.8456 & 0.676 & 243.5 & 600 & & 2806.8027  & 0.898 & 323.3 & 600 \\
  2804.8531 & 0.677 & 243.8 & 600 & & 2807.8005  & 0.011 & 4.0 & 600 \\
  2805.8207 & 0.787 & 283.4 & 600 & & 2808.7530  & 0.119 & 42.9 & 600 \\
  2805.8282 & 0.788 & 283.7 & 600 \\
  2806.8490 & 0.904 & 325.3 & 900 \\
  2807.8477 & 0.017 & 6.1 & 900 \\
  2808.8297 & 0.128 & 46.2 & 600 \\
  2808.8370 & 0.129 & 46.5 & 600 \\
  \\
  \hline
 \multicolumn{2}{c}{6506 - 6620\AA} & \multicolumn{2}{c}{R = 95,000} & & \multicolumn{2}{c}{6592 - 6702\AA} & \multicolumn{2}{c}{R = 96,000} \\
 \hline
 HJD & \multicolumn{2}{c}{Phase} & Exp. time & & HJD & \multicolumn{2}{c}{Phase} & Exp. time \\
 (2450000+) & (Decimal) & (\degr)~~& (s) & & (2450000+) & (Decimal) & (\degr)~~ & (s)  \\    
 \hline
 2801.8222  & 0.334 & 120.1 & 900 & & 2800.8628 & 0.225 & 80.9 & 900 \\
 2802.8313  & 0.448 & 161.3 & 900 & & 2801.7982 & 0.331 & 119.1 & 600 \\
 2803.8225  & 0.561 & 201.8 & 900 & & 2801.8055 & 0.332 & 119.4 & 600 \\
 2804.8245  & 0.674 & 242.7 & 600 & & 2802.8076 & 0.445 & 160.4 & 600 \\
 2805.7971  & 0.785 & 282.4 & 600 & & 2802.8148 & 0.446 & 160.6 & 600 \\
 2806.7459  & 0.892 & 321.1 & 900 & & 2803.7967 & 0.558 & 200.8 & 600 \\
 2807.7535  & 0.006 & 2.3 & 600 & & 2803.8043 & 0.559 & 201.1 & 600 \\
 2808.8151  & 0.127 & 45.6 & 600 & & 2804.8019 & 0.672 & 241.7 & 600 \\
  & & & &                                & 2804.8094 & 0.673 & 242.1 & 600 \\
  & & & &                                & 2805.7744 & 0.782 & 281.5 & 600 \\
  & & & &                                & 2805.7820 & 0.783 & 281.8 & 600 \\
  & & & &                                & 2806.7644 & 0.894 & 321.9 & 600 \\
  & & & &                                & 2806.7719 & 0.895 & 322.2 & 600 \\
  & & & &                                & 2807.7725 & 0.008 & 3.0 & 900 \\
  & & & &                                & 2808.7929 & 0.124 & 44.7 & 600 \\
  & & & &                                & 2808.8006 & 0.125 & 45.0 & 600 \\
  \hline
 \end{tabular}
 }
\end{table*}

\section{Spectroscopic data}

Spectra of a\,Cen were obtained with the Canada-France-Hawaii Telescope (CFHT) and 
the high-resolution Gecko coude spectrograph on nine consecutive nights in 2003 June.
Since Gecko's mosaic grating is a single-order echellette, the spectra were obtained at 
four different grating rotations to obtain 75 to 125\AA-long spectra in 
regions which included the \ion{He}{i} lines $\lambda\lambda$4437, 4471, 6678 and 
7281, as well as H$\alpha$.
The \ion{He}{i} $\lambda$7281 line has rarely been observed because of strong 
atmospheric contamination in this spectral region but its relative 
weakness, large isotope splitting, and the very high revolving power provided by Gecko in
this order ($R = 158,000$) makes it an ideal diagnostic for \element[][3]{He}.

The spectra were processed with IRAF\footnote{IRAF is distributed by the 
National Optical Astronomy
Observatory, which is operated by the Association of Universities for
Research in Astronomy (AURA), Inc., under cooperative agreement with the
National Science Foundation.} and included telluric line removal in the 
H$\alpha$ and \ion{He}{i} $\lambda$7281 spectral regions using the bright, 
rapidly rotating star $\alpha$~Leo as the telluric standard.
Typical continuum S/N of the individual reduced spectra is about 500 (300 for the
$\lambda$7281 spectral region) and the 
resolving power, as determined by the FWHM of the comparison arc lines obtained before and after each grating rotation change, 
ranges from $93,000 < R < 158,000$. 

Phases for each observation were obtained using the most recent ephemeris 
determined for a\,Cen \citep{catalano96}:

\begin{equation}
JD (u_{min}) = 2,442,808.376 + 8.817744 \pm 0.000109\times E.
\end{equation}

\begin{figure}[!t]
\resizebox{\hsize}{!}{\includegraphics{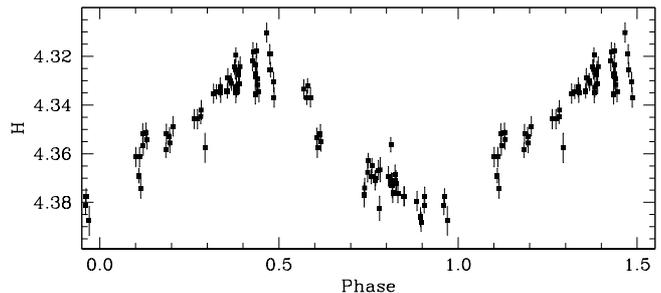}}
\vspace{-5cm}
\caption{Hipparcos photometry for a\,Cen plotted using the ephemeris given in the text.
The photometric minimum in $H$ appears to be shifted by 0.05 to 0.1 cycles relative to this
ephemeris defined by the $u$ light curve.}
\label{hipparcos}
\end{figure}

\noindent They improved the precision of the original period determined by \cite{norris68} by combining data from \citet{norris71} and \citet{pedersen77} with new $uvby$ photometry.
While this ephemeris appears to combine data together to produce a well-defined 
minimum at $\phi=0$ in $u$, the helium line strength index $R$ shown in their Figure~3 
appears to indicate that the helium line strength maximum, as well as the $vby$ photometry,
precedes the $u$ photometric minimum by a small phase increment of perhaps 0.05 to 0.10.  
As shown below, this is significant in that the helium maximum for our spectroscopic 
data appears to precede the zero-phase calculated from the above photometric 
ephemeris as well. 
In Fig.~\ref{hipparcos} we show that Hipparcos photometry of a\,Cen also shows the same
phase shift.
This difference in phase appears to be real since shifting the Hipparcos photometry to give a minimum at phase zero would require a reduction of the period by an amount approximately
ten times its quoted uncertainty.
There has only been about 580 and 1130 stellar rotations between the \citet{catalano96}
observations, the Hipparcos observations and our
spectra acquisition so the more recent data should only have errors of about 0.007 and 0.014
in phase respectively. 
Based on this discussion we will therefore refer to the ephemeris above 
as identifying the $u$ minimum. 

As noted above, the negative extremum of a\,Cen's magnetic field occurs near the phase of maximum helium line strength or photometric minimum.


\begin{figure}[!t]
\resizebox{\hsize}{!}{\includegraphics{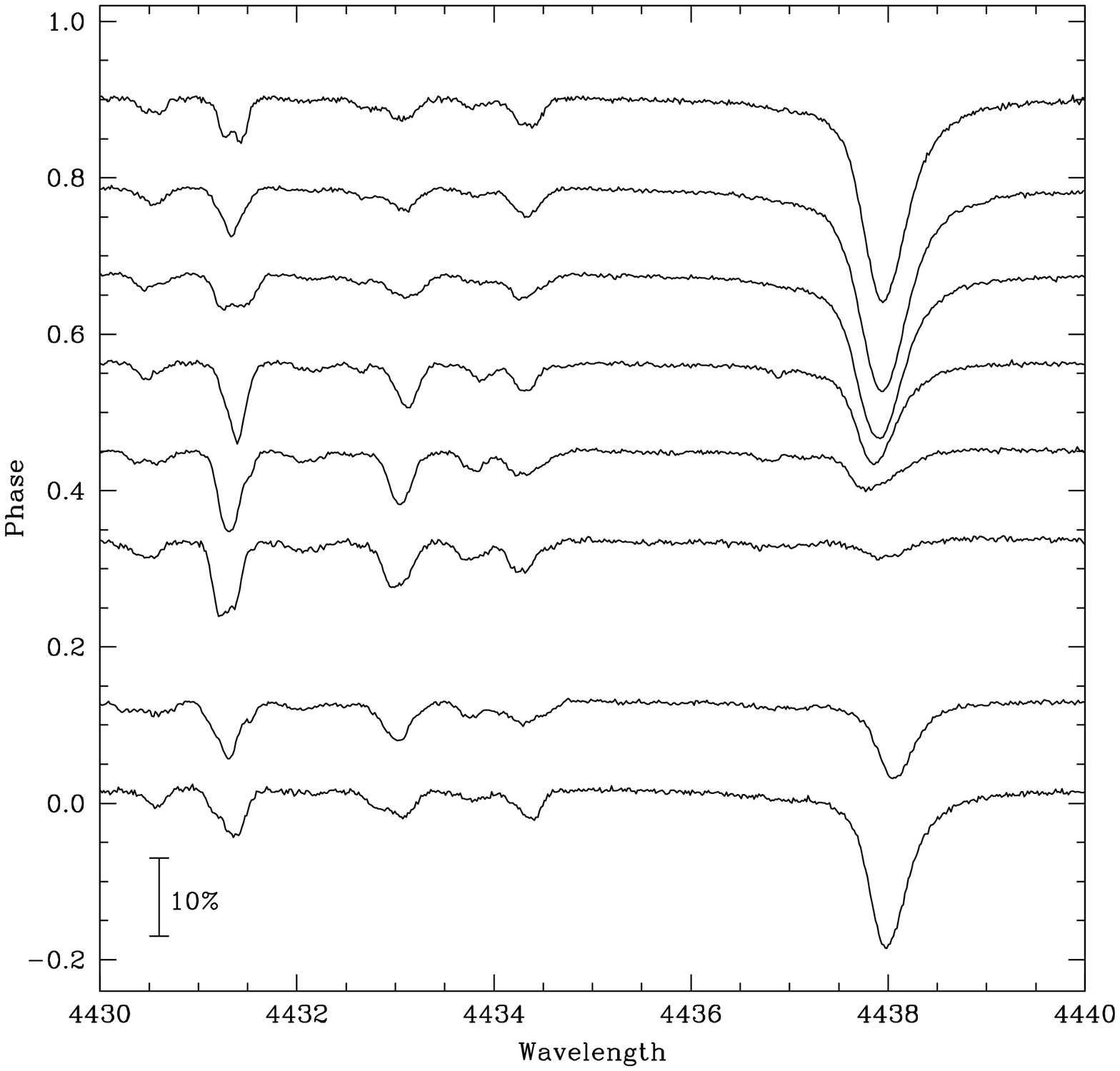}}
\caption{A small sample of the spectra of a\,Cen discussed in the paper.
The continuum of each spectrum is plotted at a height corresponding to the phase of the observation calculated using the ephemeris given in the text.
The relative intensity scale is shown by the vertical bar at the lower left. 
Note the extreme change in the strength of the \ion{He}{i} $\lambda$4437 line 
as well as the pronounced profile shape variability of the metal lines.}
\label{sample_spec}
\end{figure}

Table~\ref{slog} provides a summary of our spectroscopic observations.  
Rotational phases are provided in both decimal and angular 
formats for easy comparison with our models below.  
A small sample of the spectra in a section of a single spectral window are 
presented in Figure~\ref{sample_spec}.


The number of observations and the uniformity of phase coverage is much less than 
ideal for Doppler imaging.
However the high-resolution of the data and the surprising degree of profile
shape variations apparent in Figure~\ref{sample_spec} invites an attempt to extract 
some information about the
distribution of elements such as He, Fe, N and O over the surface in the hope that
the symmetry of distribution will augment what is known about the magnetic
variability and perhaps indicate the orientation of the magnetic axis 
relative to the rotational axis.

\section{Doppler imaging}

The technique for Doppler imaging of Ap and Bp stars to obtain a map of the variable 
abundance of elements over their surfaces has been thoroughly covered in the
literature. Examples of the computing approaches to Doppler imaging are reviewed in 
\citet{back1} and in \citet{back2} and a comprehensive review
of many of the uncertainties in Doppler imaging, although focussing in this case
mainly on cool stars, is found in \citet{back3}. 
 
A critical issue for Doppler imaging is whether the same $T_{\rm eff}$~can be assumed 
for the entire surface of the star or whether a variable $T_{\rm eff}$~is required in 
the programming.  
Fortunately this issue has been reviewed in an extensive and careful analysis 
carried out by \citet{molnar74} based on the OAO-2 ultraviolet photometry scans. 
Molnar concluded that neither the photometric variability nor the ultraviolet Si 
line variations were due to a variation in $T_{\rm eff}$ over the rotation cycle of 
the star. 
\citet{krticka07} have also demonstrated that the light curve of the helium-strong star HD\,37776, an object quite comparable to a\,Cen in fundamental properties and in having quite
pronounced line profile variations, can be almost completely accounted for by the 
inhomogeneous surface distribution of silicon and helium over the star's surface and the
bound-free transitions of these elements and not surface temperature variations.
More recent work on a cooler Ap star, HR\,7224, also suggests that surface temperature
variations do not exist in Ap stars \citep{krticka09}.

The $T_{\rm eff}$~and log~g values assumed for a\,Cen by previous authors 
\citep[e.g.][]{molnar74,norris71}  are in the range of 19,000~K to 20,000~K 
and $\log{g}$ of about 4.0.
We have used $T_{\rm eff} = 19,000$ and $\log{g} = 4.0$ for all models discussed in this paper.

A discussion of the adopted rotation axis inclination, $i$, is deferred until the end of this section.

\subsection{Helium}

Ideally, for a helium variable such as a\,Cen, it would be most informative if a Doppler image of
the star's helium distribution could be obtained with at least moderate resolution. 
There are four \ion{He}{i} lines in the four spectroscopic regions given in Table~\ref{slog}: 
the three singlet lines $\lambda\lambda$4437.551, 6678.154, and 7281.349, all arising from the same
2p\,$^1$P lower level, and the triplet $\lambda$4471 line and its forbidden component.  
The latter line is so strong and complex as to be unusable for Doppler
imaging, given the relatively slow rotation of a\,Cen. 
As a first very simple step we therefore used the 4437\AA~ line
to obtain an approximate map of the helium distribution.   
We ignored the complex Stark broadening inherent to the line and instead 
used highly adjusted values for radiative, Stark and van der Waals broadening parameters of
6.0, -3.8 and -6.5 respectively.  Wavelengths and $\log{gf}$ values adopted for the helium lines
\citep{kurucz95} are provided in Table~\ref{lines}. 

\begin{figure}[!t]
\resizebox{\hsize}{!}{\includegraphics{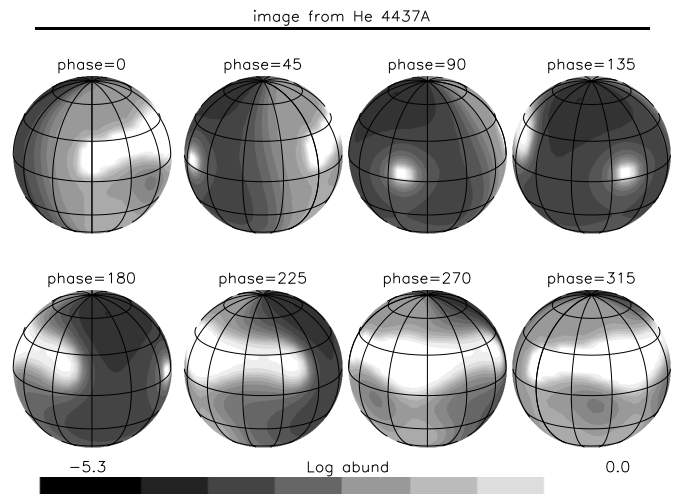}}
\caption{ An approximate map of the location of the helium concentration on a\,Cen
obtained by a fit to the $\lambda$4437 helium line.}
\label{helium}
\end{figure}

\begin{figure}[!b]
\resizebox{\hsize}{!}{\includegraphics[angle=-90]{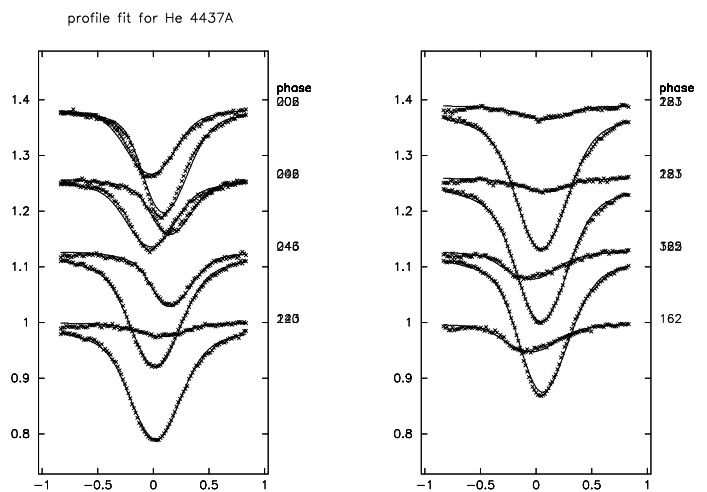}}
\caption{\ion{He}{1} $\lambda$4437 line profiles variations of a\,Cen and the results of Doppler
imaging.  Observed profiles are represented by the points while the model fits are solid lines.
The phases of the observed profiles are marked in degrees on the right side of the figure and
the wavelength relative to line centre (in \AA) is indicated at the bottom.}
\label{helium_DIP}
\end{figure}

We show our Doppler image of the distribution of helium over the surface of a\,Cen based on
the $\lambda$4437 line in Fig.~\ref{helium}.
Where multiple consecutive spectra were obtained we did not combine spectra; instead this
provided an indication of the noise level of the spectra in the plots of line profile fits that follow.
The fit to the line profiles shown in Fig.~\ref{helium_DIP} is quite good and the abundance geometry is consistent with the single spot models proposed by \citet{norris72} and \citet{mihalas73}: one hemisphere of the star has an enhanced abundance of helium while the other hemisphere is very helium deficient. 
Table~\ref{lines} gives the approximate range of helium abundances determined by the Doppler imaging.
These are averaged over several adjacent pixels but given the rather poor time sampling should not be interpreted too literally.   When abundances are averaged over the much larger obviously helium-rich and helium-difficient regions on opposite hemipheres of the star our Doppler imaging suggests that the helium abundance likely varies by a factor of 100 to 150 from one side of the star to the other. 

Interestingly, we discovered some evident problems when working with the other helium lines of 
a\,Cen to produce a Doppler image of the helium surface distribution.  
These appear to go beyond what one might expect to have because of our neglect of Stark
broadening, non-LTE affects,  possible element stratification or simply a poor choice of line. 
In trying to invert the profiles of the $\lambda$6678 line to form an
approximate map of the helium distribution it was found that the program required 
an assumption of a stellar radial velocity that was 5 km\,s$^{-1}$ more positive
than was needed for the lines of the other elements as discussed in the next section. 
This seemed to arise because
of a distortion redward of the center of the line profiles that appears roughly between 
phases 100\degr\ and 200\degr. 

\begin{figure}[!t]
\resizebox{\hsize}{!}{\includegraphics[angle=-90]{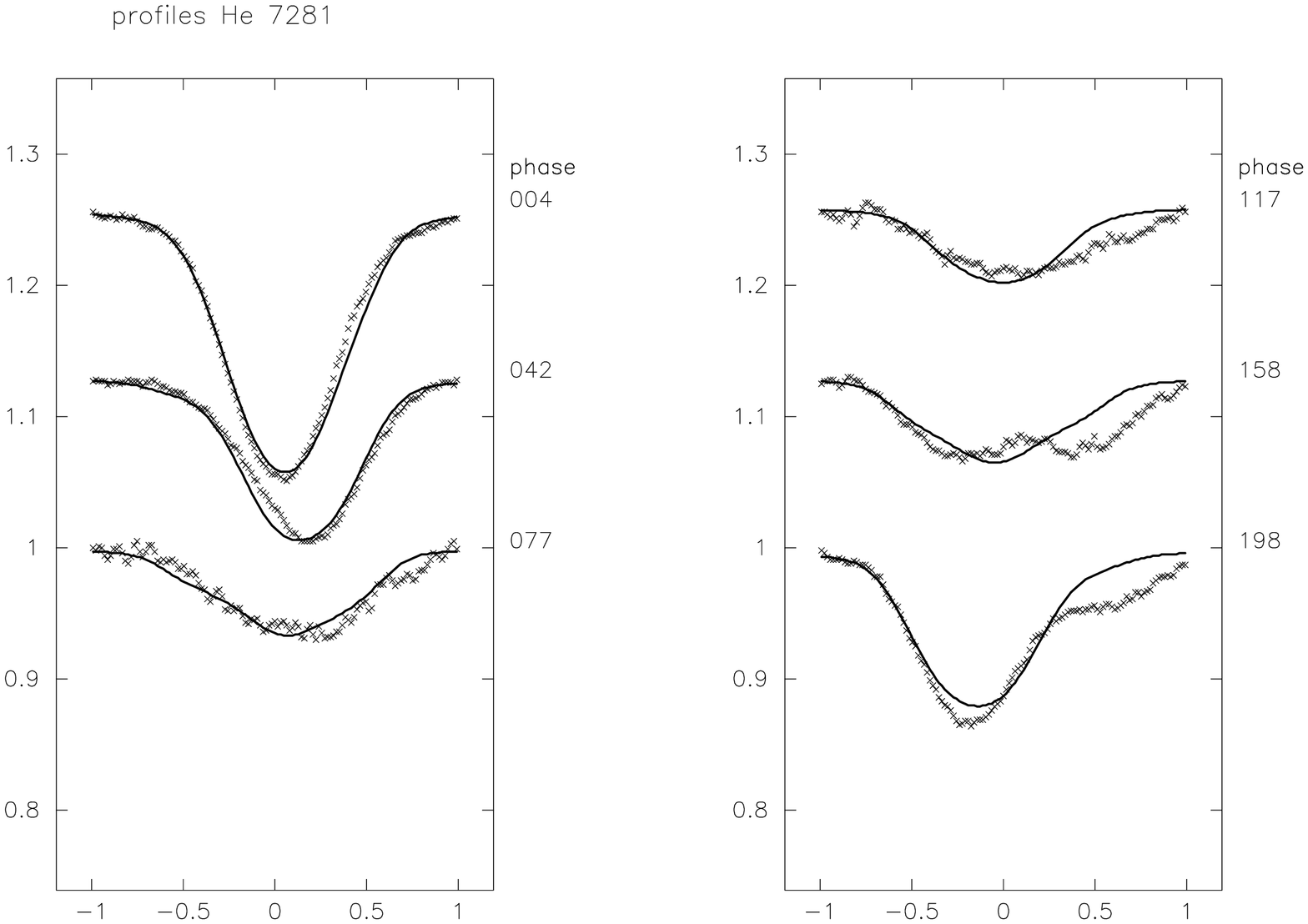}}
\resizebox{\hsize}{!}{\includegraphics[angle=-90]{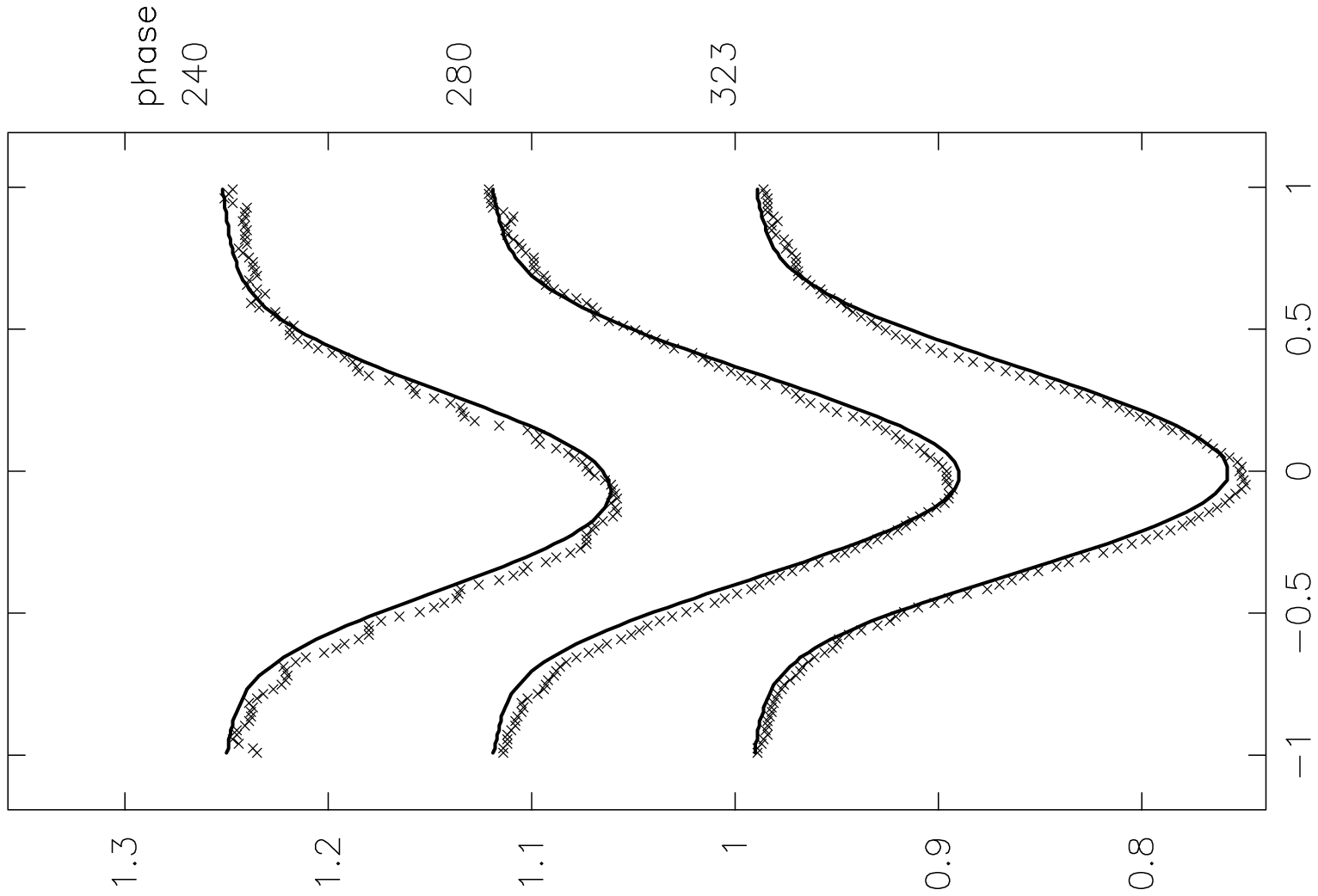}}
\caption{As for Fig.~\ref{helium_DIP} but for the \ion{He}{1} $\lambda$7281 line profile.
Conventional Doppler Imaging is not able to reproduce the ``satellite'' feature longward of the
line center between phases of 100\degr\ and 200\degr.}
\label{helium_DIP_7281}
\end{figure}

Attempts to reconstruct a helium abundance map using
the $\lambda$7281 line showed a similar problem as the $\lambda$6678 line,
but even more exaggerated.
The Doppler image profile fits that should match
the observed profiles at phases up to about phase 100\degr\ are confounded by a strong
satellite feature about 0.5\AA\ to the red that persists from about phase 100\degr\
to about phase 200\degr\ and then quickly and fairly completely disappears for the rest
of the stellar rotation. The attempts by the program to accommodate this anomalous feature
lead to persistent red-shifted distortions of the fitted profiles at other phases
(see Fig.~\ref{helium_DIP_7281}).  

There are no known blends from lines of other elements at
this wavelength but a possible solution to the poor models for the $\lambda$6678 and $\lambda$7281 lines
is presented in Section \ref{heRevisited}.

\subsection{Oxygen, Nitrogen and Iron}

The lines of ionized oxygen and nitrogen as well as twice ionized iron
listed in Table~\ref{lines} were found to be the most suitable for Doppler
imaging from among those available in the two observed spectroscopic regions.
They presented the least problem in terms of blending issues, had an adjacent
continuum that was not significantly distorted by crowding lines or the presence
of broad wings of helium or hydrogen lines and were
sufficiently strong to provide good line signal-to-noise. The program used for
Doppler abundance imaging of Ap stars allows for blends in the calculation of
the local line profiles but mapping of the abundance distribution over the
stellar surface is only done for one element at a time so blends must either be
of a non-variable element or blends of lines of the same element \citep[see for
example the mapping of the 7775\AA\ line triplet of oxygen for $\epsilon$ UMa in]
[]{rwh}. The data for the lines listed in Table~\ref{lines}   
come from the VALD database \citep{vald}.

\begin{table}[t]
\caption{Lines used for Doppler imaging of a\,Cen and the approximate range of elemental abundances
over the surface of the star.}
\label{lines}
\begin{center}
 \begin{tabular}{cccrcc}
  \hline
 Wavelength & Ion & E.P. & $\log{gf}$ & Min. & Max. \\
 (\AA)  & & (eV) &  & $\log{N/N_H}$ & $\log{N/N_H}$ \\
   \hline   
  4437.551 & He I & 21.242 & -2.034 & -3.9 & -0.2 \\
  6678.154 & He I & 21.242 & 0.329 \\
  7281.349 & He I & 21.242 & -0.842 \\
  4432.736 & N II   & 23.415 &  0.580 & -4.9 & -2.9 \\
  4447.030 & N II   & 20.409 &  0.285  \\
  6610.562 & N II   & 21.600 &  0.433  \\
  4414.900 & O II   & 23.442 &  0.172 & -4.3 & -1.4 \\
  4416.980 & O II   & 23.419 & -0.077  \\
  4419.596 & Fe III &  8.241 & -2.218 & -5.8 & -2.0 \\
  4431.019 & Fe III &  8.248 & -2.572  \\
  \hline
 \end{tabular}
\end{center}
\end{table}

Doppler images were generated using the lines of Table~\ref{lines} for the
elements iron, nitrogen and oxygen. The images were intercompared 
with one another and then compared with the general symmetry discussed by 
\citet{mihalas73} and as suggested by the helium distribution as generally 
represented in Fig.~\ref{helium}.

\begin{figure}[!t]
\resizebox{\hsize}{!}{\includegraphics{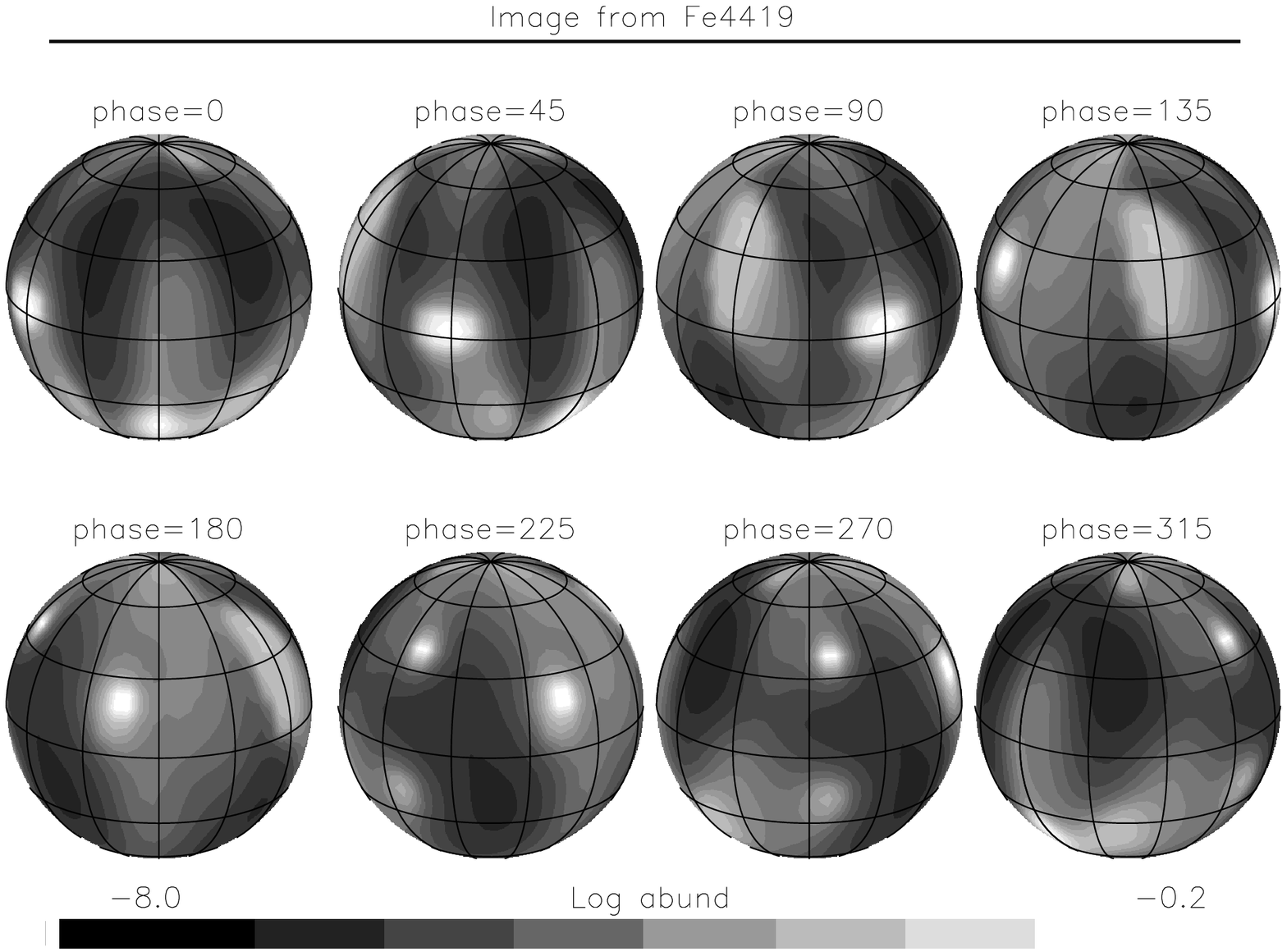}}\vspace{4mm}
\resizebox{\hsize}{!}{\includegraphics{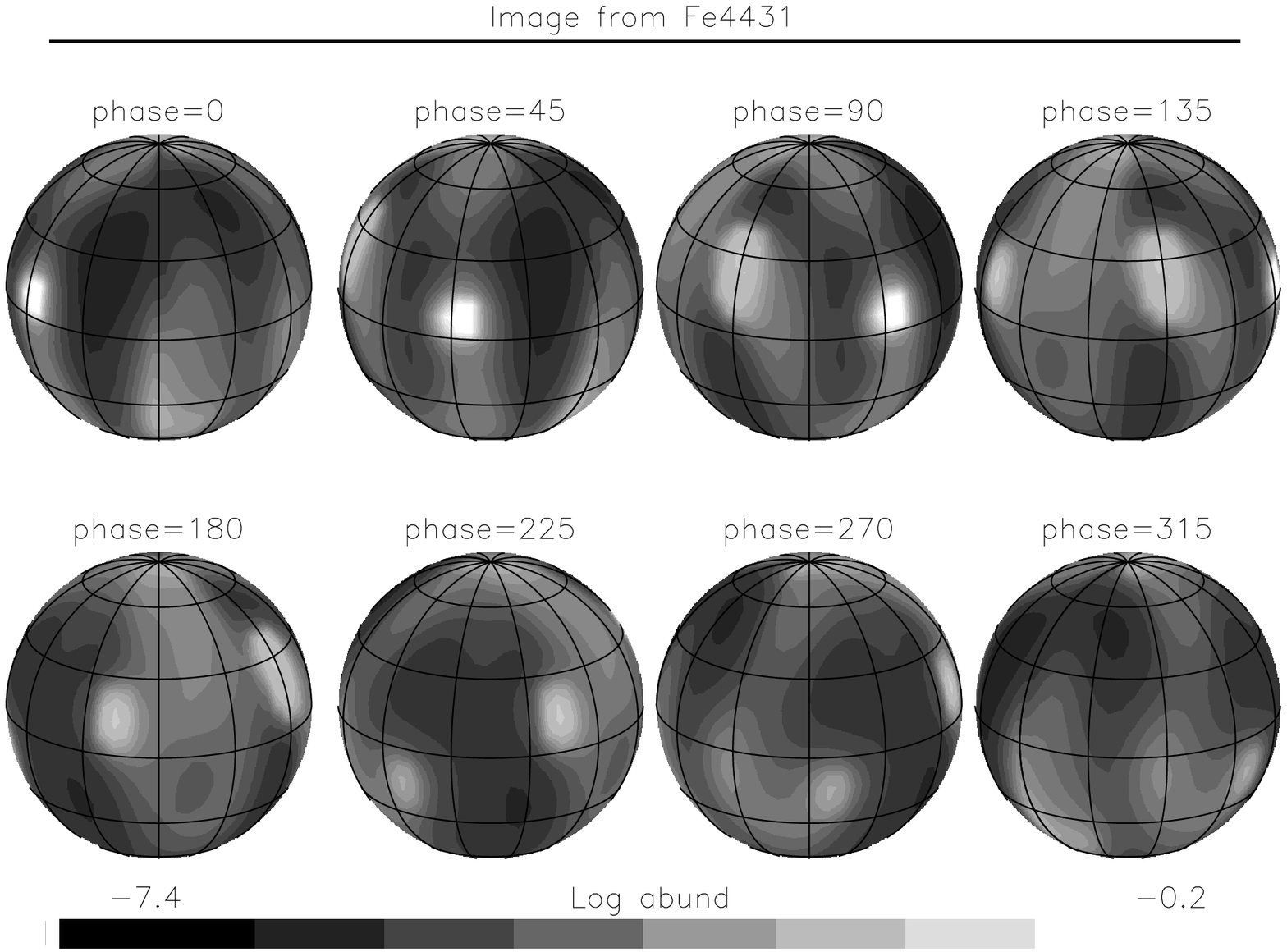}}\vspace{4mm}
\resizebox{\hsize}{!}{\includegraphics{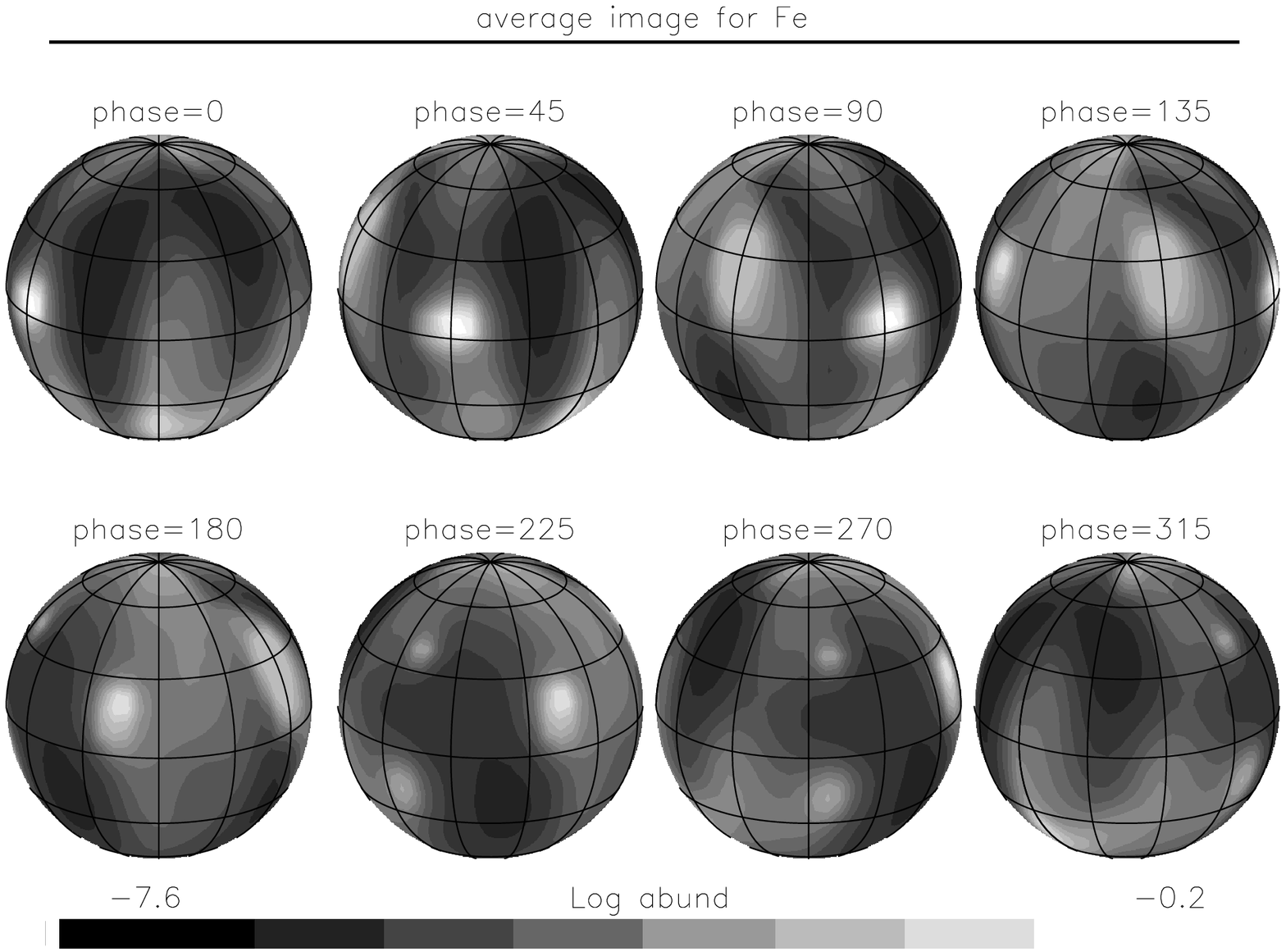}}
\caption{ Doppler images of the distribution of Fe on the surface of a\,Cen reconstructed
from the profile variations of the 4419\AA\ and 4431\AA\ lines separately, and an image created
from the combination of the Doppler images of both features.}
\label{feFits}
\end{figure}

\begin{figure}[!t]
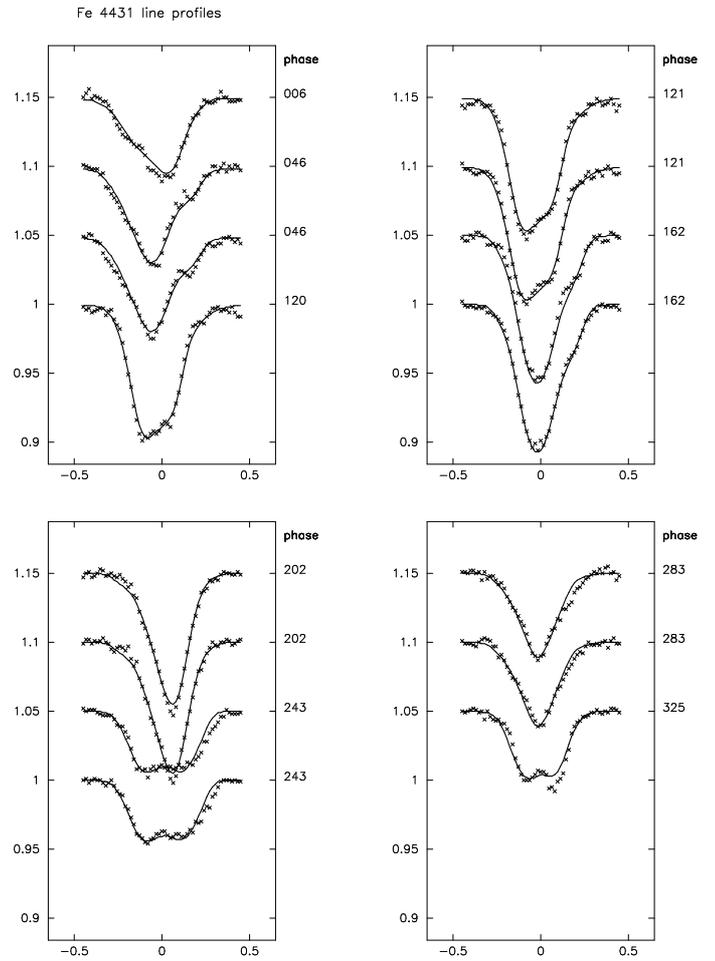

\resizebox{\hsize}{!}{\includegraphics[angle=-90]{14157fig7a.eps}}\vspace{0mm}
\resizebox{\hsize}{!}{\includegraphics[angle=-90]{14157fig7b.eps}}
\caption{As in Fig.~\ref{helium_DIP} but for the line profile variations of the iron line 4431\AA.}
\label{profiles}
\end{figure}

The images developed from the Fe lines at 4419\AA\ and 4431\AA\ are shown as 
Fig.~\ref{feFits}. The observed and model fitting to the 
line profiles for the 4431\AA\ iron line are shown in Fig.~\ref{profiles}. The
fitting to the line profiles for Fe 4419\AA\ are very similar. The regions of
enhanced iron evident on the Doppler images obtained using the two Fe lines
are well defined and relatively tightly confined in area such that the five 
regions of peak iron abundance produce clearly identifiable dips or 
distortions migrating through the line profiles with the phase of the rotation 
of a\,Cen. Note the very strong similarity of the two images obtained using 
these two lines independently.

\begin{figure}[b]
\resizebox{\hsize}{!}{\includegraphics{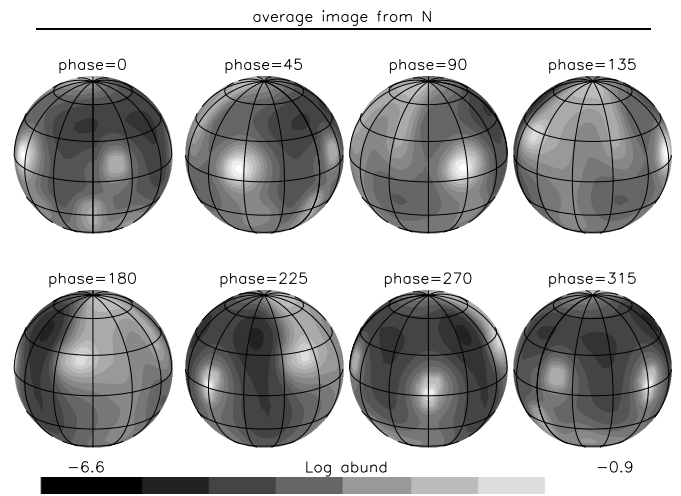}}
\caption{ A composite Doppler image of the distribution of nitrogen as constructed by
averaging the maps reconstructed through inversion of the line profiles of the nitrogen
lines listed in Table~\ref{lines}.  }
\label{nitrogen}
\end{figure}
 
The three nitrogen lines of Table~\ref{lines} were each used to produce separate
Doppler images using the same parameters as with the iron lines and as shown in
Table~\ref{param}. The line profiles for the three nitrogen lines did not show such
clear and distinct distortions passing through the line profiles with rotational 
phase as had the two iron lines. The Doppler images from each nitrogen line then did not
show such compact regions of abundance enhancement. Fig.~\ref{nitrogen} shows the
average of the Doppler images from the three nitrogen lines and it is evident that
the image obtained for the nitrogen abundance distribution is similar to that for
iron in that the placement of the abundance peaks is close to the positioning of the
iron peak abundance patches but as is suggested by the less distinct pattern of line
profile distortions the abundance patches are less extreme and broader in area. 

\begin{table}[b]
\caption{Adopted parameters for a\,Cen (HR 5378).}
\label{param}
\begin{center}
 \begin{tabular}{cc}
  \hline
  Parameter & Adopted Value\\
  \hline   
  $T_{eff}$                & 19000 K \\
  $\log{g}$                 & 4.0 \\
  $v\sin{i}$                & 15.0 km\,s$^{-1}$ \\
  Inclination $i$          & 70\degr \\
  Rotation period          & 8.817744 days \\
  Radial velocity       & $5.8 \pm 0.2$ km\,s$^{-1}$\\
  Micro turbulence $\xi$   & 2.5 km\,s$^{-1}$\\
  Macro turbulence         & 0.0 km\,s$^{-1}$\\
  Radius                   & 2.8 solar\\
  \hline
 \end{tabular}
\end{center}
\end{table}

Similar patterns of abundance distribution were obtained from the oxygen 
lines and, as with the nitrogen distributions, the pattern was generally less distinct
on the surface of the star and  the line profiles did not show the clear pattern of
migrating dips through the line profiles with rotational phase. (Although, as
with nitrogen, the line profiles exhibited profile distortions that changed with
phase and that were suggestive of the migrating profile distortions seen for the
iron lines). 
The approximate range of abundances determined by the Doppler imaging for each element are given in Table~\ref{lines}.

The geometric parameters used for Doppler imaging a\,Cen are given in
Table~\ref{param}. 
The choice of 70\degr\ for the inclination was more convoluted than in other cases of Doppler imaging.
The usual preliminary search for a minimum in the error of fit of the profiles calculated from the best image formed
at each of a range of inclinations to the observed profile data gave the usual broad curve with a minimum
error for the iron, nitrogen and oxygen images at smaller values of inclination around 45\degr. 
As is also usual, the image formed from the line profiles for these elements was not highly sensitive to 
inclination. 
However, the helium profiles were far better
fit with an image formed using an inclination of approximately 70\degr. 
Exploration of the images produced and
the profile fits for the Fe, N and O data showed that while the overall fit to the profiles was slightly better
at smaller inclinations, largely due to the better fit to the very bottom of the deeper profiles, the inverted
bottom of the Fe profiles in particular at the two profiles at phase 243\degr\ was far better represented in the images formed
with inclination close to 70\degr. That distinctive and well-defined feature was quite poorly fit at smaller
inclinations.

The choice of microturbulent
velocity was made by applying the Blackwell method \citep{blackwell79} embedded 
in the \emph{SPECTRUM} program \citep{gray94}. 
The program was used with the nitrogen lines
and suggested a microturbulence of 2.5 km\,s$^{-1}$.

\section{Helium Revisited}\label{heRevisited}

\citet{bohlender05} has used data identical to those presented here for a\,Cen to study the 
abundance of \element[][3]{He} in a number of well-known helium-3 stars \citep{hartoog79}.
In a\,Cen the peculiar red satellite feature in the $\lambda$7281 line, apparent
between phases 77\degr\ and 240\degr, is shifted by approximately 0.5\AA\ from the line center of the
latter line.
This is intriguingly close to the observed shift of 0.55\AA\ between the \element[][3]{He} and \element[][4]{He}
components of this line \citep{fred51}.
A similar shift is suggested for the $\lambda$6678 line, consistent 
with its isotope shift of 0.50\AA, but this feature is less obvious because of the greater line 
strength.
We have therefore tried to fit the helium lines by adding a contribution from \element[][3]{He} as 
described below.

First, we generated a grid of relatively simple pure hydrogen and helium non-LTE model atmospheres with $T_{\rm eff} = 19,000$\,K, 
$\log{g} =4.0$ and a range of $N$(He/H) from 0.001 to 0.8.  
We used the \emph{TLUSTY} program \citep{hubeny95} and started with the 
published \citep{lanz07} line-blanketed model atmosphere BG19000g400v2 
with $T_{\rm eff} = 19,000$\,K, $\log{g}=4.0$ 
and $N$(He/H)$=0.1$ as the input atmosphere for the program.
The companion spectral line synthesis program \emph{SYNSPEC} was then used to produce
grids of specific line and continuum intensities for the combined \element[][3]{He} and 
\element[][4]{He} lines for each helium abundance.
To do so we first had to modify the version of \emph{SYNSPEC} we used (v48 provided by
Lanz, private communication) to incorporate broadening parameters of \citet{dimitrijevic90} 
for the $\lambda$7281 line since these were not included in the source code.
\emph{SYNSPEC} was also modified so that it would recognize the wavelengths of the \element[][3]{He} components of line profiles and use the same broadening parameters as are used for the corresponding \element[][4]{He} lines. 

Unfortunately \emph{TLUSTY} does not permit incorporation of separate \element[][3]{He} and 
\element[][4]{He} abundances in the production of a model atmosphere.  
We have therefore assumed that the helium abundance used
for the \emph{TLUSTY} models is the combined \element[][3]{He} $+$ \element[][4]{He} 
abundance and also that the lower mass of the lighter isotope has no affect on the structure 
of the atmosphere, or on the treatment of broadening for the lines.
We then included both \element[][3]{He} and \element[][4]{He} line data in the input line list 
used by \emph{SYNSPEC} and artificially adjusted the $\log{gf}$ 
values of the lines to set the relative abundances of the two isotopes but keep the total helium abundance used by
\emph{SYNSPEC} consistent with the abundance used in the model atmosphere.   
For example, in the case of the $\lambda$7281 line if 
$N$(\element[][3]{He} + \element[][4]{He}$) = 0.1$ in the \emph{TLUSTY}
model and $N$(\element[][3]{He}/\element[][4]{He}$)=1$ then identical $\log{gf}$ values 
of $(-0.842 - \log{2}) = -1.143$ were used for both isotopes in the input line list for \emph{SYNSPEC}
to provide $N$(\element[][3]{He}/H) $= 0.05$ and
$N$(\element[][4]{He}/H) $= 0.05$.
(The $\log{gf}$ value for the \element[][4]{He} 
$\lambda$7281 line is $-0.842$.)

The specific intensities were then used in a disk integration program that permits the placement of circular 
zones of 
different helium abundances anywhere on the disk to approximately model the surface abundance geometry.  
Each zone is defined by an angular radius and its colatitude and longitude relative to a point that crosses 
the line of sight to the observer
at $\phi = 0$ \citep[see Figure 3 of][for an illustration]{bohlender90}.  
Bands of different abundances can also be modeled with the appropriate superposition of two circular patches.
At each point of the disk integration the local helium abundance is determined and specific intensities for 
the appropriate
line and local model atmosphere are used to construct the line profile.

\begin{figure}[!t]
\resizebox{\hsize}{!}{\includegraphics[angle=-90]{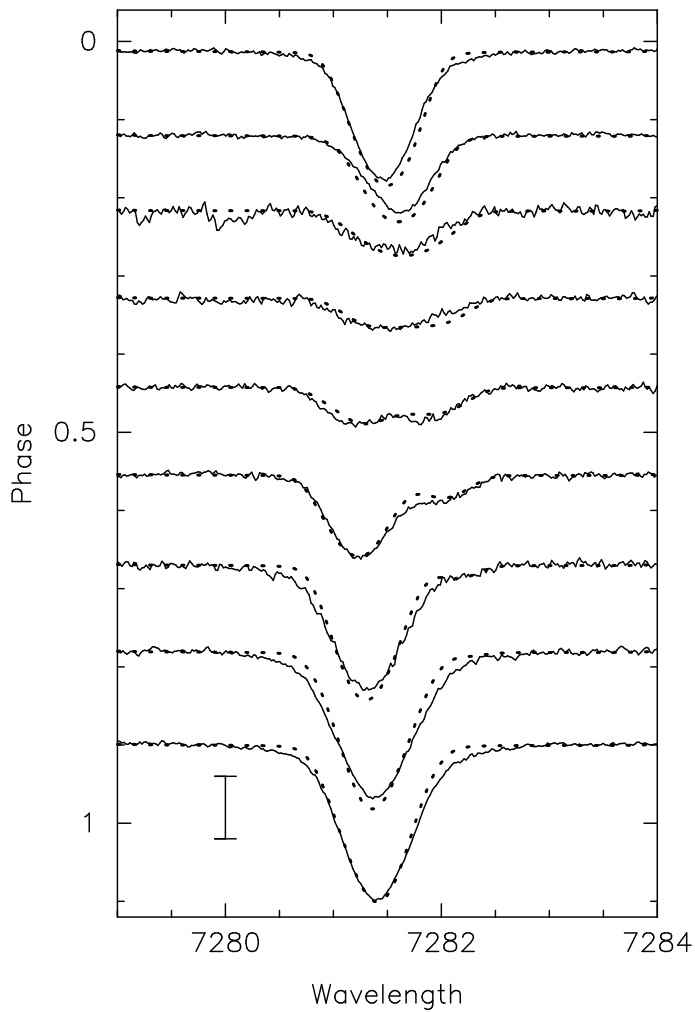}}
\caption{Model fits for the \ion{He}{i} $\lambda$7281 line of a\,Cen in which contributions
from both \element[][3]{He} and \element[][4]{He} isotopes are incorporated.  
The continuum of each spectrum is plotted at a height corresponding to the phase of the observation calculated using the ephemeris given in the text.
The relatively intensity scale is shown by the vertical bar at the lower left.
The helium deficient hemisphere of a\,Cen appears to have a huge overabundance of \element[][3]{He}.  }\label{helium3Fit}
\label{helium3}
\end{figure}

Line profiles were generated for the \ion{He}{i} $\lambda\lambda$4437, 4471, 6678 and 7281 lines.  
Fitting was conducted on a trial and error basis until reasonable fits were found for each line for as
simple a configuration of abundance spots as possible.
Our models indicated that a large inclination of about 70\degr\ was required to reproduce the line
profiles and it was this result that led us to revisit the inclination derived from Dopper imaging of the metal
lines as described in the previous section.
Figure \ref{helium3Fit} shows the resulting fits to the $\lambda$7281 helium line profiles as
a function of phase.
While a model with just two roughly hemispherical regions with widely differing helium abundance produced a reasonably 
good fit, we were able to produce a substantially better fit to the line and the red satellite feature
at all rotational phases by using three regions of different, but uniform, abundances.
The illustrated fit is produced with a 55\degr\ wide helium-rich band with $N($\element[][4]{He}/H$) = 0.50$
centered at a colatitude of 300\degr\ and extending from 25\degr\ to 80\degr\ from what is assumed to be 
the negative magnetic pole.
The 25\degr\ radius polar region centered at the same colatitude has an approximately solar helium abundance of $N($\element[][4]{He}/H$) = 0.10$
The rest of the star's surface has
$N(($\element[][3]{He}+\element[][4]{He})/H$) = 0.004$ and 
$N($\element[][3]{He}/\element[][4]{He}$) = 1$.
In other words, a broad helium-rich region covers much of one hemisphere of the star, except for small region with
a normal abundance near the pole, while the other hemisphere is
extremely helium poor but has a huge relative overabundance of \element[][3]{He}.
The \element[][3]{He} contribution to the line profile is readily apparent between phases 0.3
and 0.7.
There is not a discernible contribution to the helium line profiles from \element[][3]{He} in the helium-rich
hemisphere.
The broad helium-rich band's colatitude of 300\degr\ indicates that it crosses the line of site to the 
observer at approximately the same phase as the Hipparcos photometric minimum discussed
earlier and is also consistent with the phasing of the helium line strength photometry.

Good fits to the \ion{He}{i} $\lambda\lambda$4437, 4471 and 6678 lines can be produced with
very similar models. 
Minor adjustments to the helium abundances are required to improve the fits to these lines but
this is not surprising.
\citet{bohlender05} has shown that in a few of the previously recognized helium-3 stars both isotopes of helium show evidence of abundance stratification, with \element[][3]{He} located higher in the atmosphere than \element[][4]{He}.
Our unstratified models of the $\lambda$6678 line and especially the $\lambda$4437 line require somewhat lower 
\element[][4]{He} abundances in the helium-rich regions as well as lower \element[][3]{He} abundances in the 
helium-weak hemisphere to produce satisfactory models.  This is exactly what is seen in other helium-3 stars so we suggest,
as did \citet{leone97},
that both isotopes of helium have a vertical abundance stratification in the atmosphere of a\,Cen.
We will not pursue this question in detail until data with better phase resolution are available.  
More significant modifications of \emph{TLUSTY} and \emph{SYNSPEC} will also be needed to accommodate different vertical stratifications of both isotopes of helium in the generation of future (preferably fully-blanketed) non-LTE
model atmospheres and line profiles.

Qualitatively, the overall helium abundance geometry agrees with our Doppler Imaging result: a\,Cen is dominated by
a helium-rich region covering much of one hemisphere of the star.  The small polar region with somewhat lower 
abundances is also obvious at phase 315\degr\ in Fig.~\ref{helium}.
However, the most remarkable result of the isotope abundance fitting is the fact that the very helium deficient hemisphere of the star has a very large relative abundance of \element[][3]{He}.


\section{Discussion}

As was discussed in the introduction to this paper, the ephemeris used here is the
photometric ephemeris of \citet{catalano96} where minimum light is
at phase 0 but from their Fig. 3 in their paper where they simultaneously show 
the light curves with the variation of the helium index R it is clear that the
helium maximum preceeds phase 0 by a significant fraction of a rotational cycle,
perhaps of the order of 0.10. The crude Doppler image of Fig.~\ref{helium} and the profile
fits shown in Fig.~\ref{helium3Fit} are
consistent with this as phase 315\degr\ in the image appears to be roughly when
the maximum helium line strength occurs and the center of the helium-rich hemisphere in our
simple isotope modeling is centered at 300\degr.   The minimum line strength occurs 
at phases of between 120\degr\ and 135\degr. As conjectured by \citet{mihalas73} the concentration of
helium abundance is at about 90\degr\ to the rotational pole (i.e. at a latitude of
about 0\degr). The location of the maximum of helium abundance appears, 
from the work of \citet{borra83}, to be close to the location of the
negative magnetic pole and given the symmetry of the magnetic variation, we may
assume the positive pole is close to latitude 0\degr\ near a longitude (phase) of
135\degr. 

We now want to compare the location of the abundance peaks that appear consistently  
in the multiple images derived from the separate lines of iron, oxygen and nitrogen.
The location of these peaks for these elements are very similar as noted above and
shown in Fig.~\ref{feFits} and \ref{nitrogen}.
In these figures we see that there are generally
five spots. A strong pair evident in the image at phase 135\degr\ and just above the
equator. This pair would appear to be close to the location of the magnetic positive
polar region and the helium minimum but significantly these are two of the three
strongest spots on the surface rather than the weakest. Further, they seem to be
slightly above the equator rather than on it and they straddle the location of the
polar region. This bifurcation of the peak abundance into two spots is clearly not
some artifact of noise in the profile data since as has been pointed out, the
individual abundance maximum regions for the iron lines are so well defined in area
that they create clear dips seen migrating through the line profiles with
phase (Fig.~\ref{profiles}). Matching this pair of abundance peaks at phase
315\degr\ we see a weaker pair of abundance peaks near the equator and straddling the
location of the helium maximum and negative magnetic polar site.

The short description of the location of four of the five obvious abundance
concentrations is that there are two stronger concentrations straddling the location of
the positive magnetic polar region where the helium is weakest and two weaker
concentrations straddling the location of the negative polar region where the helium is
concentrated. 

This leaves as a puzzle the very evident concentration near the equator at phase 45\degr\
that produces probably the strongest of the features we see migrating through the iron
line profiles. This feature seems totally unrelated to any of the geometry we could
associate with a simple dipole-like magnetic field structure with poles located as
described above or related in any way with the apparently simple asymmetric helium
distribution with its concentration near phase 315\degr. If we choose to assume that
abundance concentrations form where the magnetic field lines are perpendicular to the
surface of the star, then it would appear that a\,Cen has a more complex field structure
than just a simple dipole or dipole plus quadrupole pattern.  
The somewhat cooler helium-weak star HD\,21699 is another example of a star with
a single spot of enhanced helium abundance on an otherwise extremely helium-weak surface
\citep{brown85,shavrina10}.  It also appears to have a very peculiar magnetic field geometry \citep{glagolevskij07}.

Investigations such as that
by \citet{bagnulo99} found that while some stars can have their magnetic field
modelled adequately by a dipole or dipole plus quadrupole field, most of their sample
required a more complex field.  In the particular case of the star 53\,Cam with its
relatively slow rotation and strong magnetic field it was seen \citep{kochukhov04} 
that the field was quite complex.  A future attempt at producing a Zeeman Doppler
image of a\,Cen will be a challenge, but would be of considerable interest
since the abundance distribution of the
lighter elements examined here suggests that the magnetic field may deviate considerably from a simple
structure.
We encourage new magnetic field measurements of any kind for the the star to greatly improve our
 understanding of
the abundance and magnetic field geometry of the star and the relationship between the two.

Once more detailed surface abundance and magnetic field geometries are known it would clearly also be of interest to determine if the relatively large ($\approx$ 0.06 mag) photometric
variations observed in the star (Fig.~\ref{hipparcos}) can be reproduced solely from the non-uniform abundances of the star, and in particular the extraordinary helium variability.
As an aside, we point out that despite the very large variation in helium abundance over the surface of a\,Cen
there is no discernible variability observed in the wings of the H$\alpha$ profile of the star during a single rotation.  
This already suggests that minimal temperature or surface gravity variations are present over the surface of 
the star.
Only in the few \AA\ at the core of H$\alpha$ are profile variations of any kind seen.  These closely reflect the amplitude and shape of the variations seen in the O, N, and Fe lines presented here near the same rotation phase and therefore likely arise from small atmospheric structure variations caused by the non-uniform metal abundances rather than variations in the helium abundance, temperature or gravity.

Our discovery of \element[][3]{He} in the atmosphere of a\,Cen adds a new member to the 
small group of helium-3 stars.
We believe this work also makes it the first firmly established magnetic member of this class
of objects; 
$\alpha$\,Scl and HR\,7129 are other magnetic candidates \citep{hartoog79} that we feel require confirmation of their \element[][3]{He} content with modern data.
As we have already mentioned, \citet{hunger99} suggested that a\,Cen might represent an important transitional object among the magnetic helium-peculiar stars where the effects of fractionation of radiatively driven winds may become important.
Our discovery that the star is a member of the helium-3 class of stars confirms this idea as does the suspected vertical stratification of helium in its atmosphere reported by \citet{leone97} and supported by our data.
It is interesting to note the fact that a\,Cen has a $T_{\rm eff}$ and $\log{g}$ very close to the well-established, but non-variable and non-magnetic prototype of this class of stars, 3\,Cen\,A.   Clearly the atmospheres and winds of slowly-rotating peculiar stars in this mass range, whether magnetic or not, are finely tuned to enable the separation of helium isotopes in their
atmosphere by either mass fractionation \citep{vauclair75} or light-induced drift \citep{leblanc93}.

We have already noted the discovery of weak metallic emission lines in a\,Cen by \citet{hubrig07}; the
incidence of this phenomenon appears to be almost universal among the helium-3 stars.

\section{Conclusions}

Doppler images of the abundance distribution of the elements helium, iron, oxygen and nitrogen 
in the helium-variable star a\,Cen appear to be consistent with the general geometry of the 
abundance distribution of helium proposed by \citet{mihalas73} where the location of the abundance concentrations is roughly at the rotational equator. 
The helium concentration is centred at a longitude between 300\degr\ and 315\degr\ using the 
ephemeris of \citet{catalano96} and we are able to reproduce abundance variations most consistently
with a relatively large value of 70\degr\ for the inclination of the star. 
For the three metallic species investigated
there are two strong concentrations of abundance near the equator at longitude roughly
135\degr, consistent with the positive magnetic maximum, and two somewhat weaker
concentrations of abundance near longitude 315\degr\ on the equator where the helium is
concentrated and roughly where the negative peak of the magnetic field would be found.
Another strong concentration is found near the equator at about longitude 45\degr\ and
this is not explainable in terms of any simple symmetry with the helium abundance or the
apparent magnetic field main polar locations.
We conclude that the surface magnetic field geometry of a\,Cen is likely more complex that that of a simple
dipole or dipole plus quadrupole.

Our detailed analysis of the variations of a number of helium line profiles show that the helium abundance
of a\,Cen varies by a factor of approximately 125 from one hemisphere to the other 
if suspected vertical stratification of helium in the star's atmosphere is ignored. 
A large portion of one hemisphere is helium rich, except very close to the negative magnetic pole, and the other hemisphere is very helium deficient.  
The latter region is also found to have an extreme overabundance of 
\element[][3]{He}, with $N($\element[][3]{He}/\element[][4]{He}$) \approx 1$.
As a result, a\,Cen is a new member of the helium-3 class of stars and the first convincing 
example of a magnetic member of this small group of peculiar objects.

\begin{acknowledgements}
  JBR thanks the Natural Sciences and Engineering Research Council
of Canada for their financial support of this work.    The authors thank the referee, Franco Leone, for a careful
review of the manuscript.
      
\end{acknowledgements}


\begin{thebibliography}{}

\bibitem[Bagnulo, Landolfi \& Landi Degl'Innocenti(1999)]{bagnulo99}Bagnulo, S., Landolfi, M., \& Landi Degl'Innocenti, M. 1999, \aap, 343,
865

\bibitem[Blackwell \& Shallis(1979)]{blackwell79}Blackwell, D.E., \& Shallis, M.J. 1979, \mnras, 186, 673

\bibitem[Bohlender \& Landstreet(1990)]{bohlender90} Bohlender, D.A. \& Landstreet, J.D. 1990, \apj, 358, 274

\bibitem[Bohlender(2005)]{bohlender05} Bohlender, D.A. 2005, EAS Publ. Ser., 17, 83

\bibitem[Borra, Landstreet \& Thompson(1983)] {borra83} Borra, E.F., Landstreet, J.D., \& Thompson, I. 1983, \apjs, 53, 151

\bibitem[Brown, Shore, \& Sonneborn(1985)]{brown85} Brown, D.N., Shore, S.N., \& Sonneborn, G. 1985, \aj,90,1354

\bibitem[Catalano \& Leone(1996)]{catalano96}Catalano, F.A., \& Leone, F. 1996, \aap, 311, 230

\bibitem[Dimitrijevic \& Shal-Brechot(1990)]{dimitrijevic90}Dimitrijevic, M.S., \& Sahal-Brechot, S. 1990, \aaps, 82, 519

\bibitem[Fred et al.(1951)]{fred51}Fred, M., Tomkins, F.S., Brody, J.K., \& Hamermesh, M. 1951, Phys.~Rev., 82, 406

\bibitem[Glagolevskij \& Chuntonov(2007)]{glagolevskij07} Glagolevskij, Y.V. \& Chuntonov, G.A. 2007, Astrophysics, 50, 362

\bibitem[Gray \& Corbally(1994)]{gray94}Gray, R.O., \& Corbally, C.J. 1994, \aj, 107, 742

\bibitem[Hartoog \& Cowley(1979)]{hartoog79}Hartoog, M.R., \& Cowley, A.P. 1979, \apj, 228, 229

\bibitem[Hubeny \& Lanz(1995)]{hubeny95}Hubeny, I., \& Lanz, T. 1995, \apj, 439, 875

\bibitem[Hubrig \& Gonzalez(2007)]{hubrig07}Hubrig, S., \& Gonzalez, J.F. 2007, \aap, 466, 1083

\bibitem[Hunger \& Groote(1999)]{hunger99} Hunger, K., \& Groote, D. 1999, \aap, 351, 554

\bibitem[Jaschek \& Jaschek(1967)]{jaschek67}Jaschek, C., \& Jaschek, M. 1967, \pasp, 70, 667

\bibitem[Jaschek et al.(1968)]{jaschek68}Jaschek,C, Jaschek, M., Morgan, W.W., \& Slettebak, A. 1968, \aj, 153, 873

\bibitem[Kochukhov et al.(2004)]{kochukhov04}Kochukhov, O., Bagnulo, S., Wade, G.A., Sangalli, L., Piskunov,
N., Landstreet, J., Petit, P., \& Sigut, T.A.A. 2004, \aap, 414, 613 

\bibitem[Krti\u{c}ka et al.(2007)]{krticka07}Krti\u{c}ka, J., Mikul\'{a}\u{s}ek, Z., Zverko, J., \& Zi\u{z}\u{n}ovsk\'{y}, J. 2007, \aap, 470, 1089

\bibitem[Krti\u{c}ka et al.(2009)]{krticka09}Krti\u{c}ka, J., Mikul\'{a}\u{s}ek, Z., Henry, G.W., Zverko, J., Zi\u{z}\u{n}ovsk\'{y}, J., \& Zv\u{e}\u{r}ina, P. 200, \aap, 499, 567

\bibitem[Kupka et al.(1999)]{vald}Kupka, F., Piskunov, N.E., Ryabchikova, T.A., Stempels H.C., \& 
Weiss W.W. 1999, \aaps, 138, 119 

\bibitem[Kurucz \& Bell(1995)]{kurucz95} Kurucz, R.L., Bell, B. 1995, Atomic Line Data, Kurucz CD-ROM No. 23.

\bibitem[Lanz \& Hubeny(2007)]{lanz07}Lanz, T., \& Hubeny, I. 2007, \apjs, 169, 83

\bibitem[Leblanc \& Michaud(1993)]{leblanc93} LeBlanc, F., \& Michaud, G. 1993, \apj, 408, 251

\bibitem[Leone \& Lanzafame(1997)]{leone97} Leone, F., \& Lanzafame, A.C. 1997, \aap, 320, 893

\bibitem[Mihalas(1973)]{mihalas73}Mihalas, D. 1973, \apj, 184, 851

\bibitem[Molnar(1974)]{molnar74}Molnar, M.R. 1974, \apj, 187, 531

\bibitem[Norris(1968)]{norris68}Norris, J. 1968, \nat, 219, 1342

\bibitem[Norris(1971)]{norris71}Norris, J. 1971, \apjs, 23, 235

\bibitem[Norris \& Baschek(1972)]{norris72}Norris, J.,\& Baschek, B. 1972, \aap, 21, 385 

\bibitem[Pedersen \& Thomsen(1977)]{pedersen77}Pedersen, H., \& Thomsen, B. 1977, \aaps, 30,11

\bibitem[Piskunov \& Rice(1993)]{back2}Piskunov, N.E., \& Rice, J.B. 1993, \pasp, 105, 1415

\bibitem[Rice(1996)]{back1}Rice, J.B. 1996, in IAU Symposium 176, 
Stellar Surface Structure ed. Strassmeier \& Linsky  Kluwer, 19

\bibitem[Rice \& Strassmeier(2000)]{back3}Rice, J.B., \& Strassmeier, K.G. 2000, \aaps, 147,151

\bibitem[Rice, Wehlau, \& Holmgren(1997)]{rwh}Rice, J.B., Wehlau, W.H., \& Holmgren, D.E. 1997, \aap, 326,
988

\bibitem[Shavrina et al.(2010)]{shavrina10} Shavrina, A.V., Glagolevskij, Yu.V., Silvester, J., Chuntonov, G.A., Khalack, V.R., Pavlenko, Ya.V. 2010, \mnras, 401, 1882

\bibitem[Vauclair(1975)]{vauclair75} Vauclair, S. 1975, \aap, 45, 233

\bibitem[Wolff \& Morrison(1974)]{wolff74} Wolff, S.C., \& Morrison, N.D. 1974, \pasp, 86, 935

\end{thebibliography}
\end{document}